# Novel magnetism of Ir$^{5+}$ ions in the double perovskite Sr$_2$YIrO$_6$

G. Cao[1], T. F. Qi[1], L. Li[1], J. Terzic[1], S. J. Yuan[2,1], L. E. DeLong[1], G. Murthy[1] and R. K. Kaul[1]

[1]Department of Physics and Astronomy and Center for Advanced Materials, University of Kentucky, Lexington, KY 40506, USA

[2]Department of Physics, Shanghai University, Shanghai, China

We synthesize and study single crystals of a new double-perovskite Sr$_2$YIrO$_6$. Despite two strongly unfavorable conditions for magnetic order, namely, pentavalent Ir$^{5+}$($5d^4$) ions which are anticipated to have $J_{eff} = 0$ singlet ground states in the strong spin-orbit coupling (SOC) limit, and geometric frustration in a face centered cubic structure formed by the Ir$^{5+}$ ions, we observe this iridate to undergo a novel magnetic transition at temperatures below 1.3 K. We provide compelling experimental and theoretical evidence that the origin of magnetism is in an unusual interplay between strong non-cubic crystal fields and "intermediate-strength" SOC. Sr$_2$YIrO$_6$ provides a rare example of the failed dominance of SOC in the iridates.

The iridates have become a fertile ground for studies of new physics driven by strong spin-orbit coupling (SOC) that is comparable to the on-site Coulomb (U) and crystalline electric field interactions. This unique circumstance creates a delicate balance between interactions that drives complex magnetic and dielectric behaviors and exotic states seldom or never seen in other materials. A profound manifestation of this competition is the novel "$j_{eff}= 1/2$ Mott state" that was recently observed in the layered iridates with tetravalent $Ir^{4+}(5d^5)$ ions [1-3]. In essence, strong crystal fields split off 5d band states with $e_g$ symmetry, whereas the remaining $t_{2g}$ bands form $j_{eff} = 1/2$ and $j_{eff} = 3/2$ multiplets via strong SOI. The $j_{eff} = 3/2$ band is lower in energy and is hence fully filled, leaving the $j_{eff} = 1/2$ band which is of higher energy half filled. The key, surprising result is the $j_{eff} = 1/2$ band has a small enough bandwidth that even a modest coulomb repulsion U among the 5d-electron states is sufficient to open a Mott gap in these iridates [1, 2], which is contrary to expectations based upon the relatively large unperturbed 5d bandwidth.

A great deal of recent theoretical and experimental work has appeared in response to early experiments, including predictions of a large array of novel effects in 5d-electron systems having strong SOI: superconductivity [4], Weyl semimetals with Fermi arcs [5], topological insulators and correlated topological insulators with large gaps, Kitaev spin liquids [6-18], etc. Most of these discussions have focused on the tetravalent iridates in which the Kramers degeneracy of the $Ir^{4+}(5d^5)$ ions results in magnetism. On the other hand, very little attention has been drawn to iridates having pentavalent $Ir^{5+}(5d^4)$ ions, primarily because the strong SOC limit is expected to lead to a nonmagnetic singlet ground state, which can be simply understood as a $J_{eff} = 0$ state arising from four electrons filling the lower $j_{eff} = 3/2$ quadruplet (see **Fig. 4 (c)** for a cartoon picture). Indeed, the $J_{eff} = 0$ state has been used to explain the absence of magnetic ordering in the pentavalent post perovskite NaIrO$_3$ **[19]**.

An interesting issue that has received some but limited attention is how the $j_{eff}$ picture is affected by non-cubic crystal fields. The presence of such crystal fields has been clearly observed in a number of recent experimental works on materials with the Ir$^{4+}$(5d$^5$) electronic configuration such as Sr$_3$IrCuO$_6$ **[20]**, Na$_2$IrO$_3$ and Li$_2$IrO$_3$ **[21]**. It is generally agreed in these studies that for Ir$^{4+}$(5d$^5$) despite the presence of the non-cubic crystal field and its importance for the electronic structure, the basic $j_{eff}$ picture is a good starting point to understand the magnetism in these iridates. *In this work we show that, in contrast, for the Ir$^{5+}$(5d$^4$) ions of Sr$_2$YIrO$_6$ the strong non-cubic crystal field results in a breakdown of the J$_{eff}$ picture even as a starting point*. As illustrated in **Fig. 4 (c)**, the conventional strong spin-orbit coupling picture, popular in the description of the Ir$^{4+}$(5d$^5$) systems, would result in the ionic state of a single Ir$^{5+}$(5d$^4$) being a non-magnetic singlet, predicting then that Sr$_2$YIrO$_6$ should be a band insulator with no magnetism. Unexpectedly, we find instead from experiment that Sr$_2$YIrO$_6$ hosts well-formed magnetic moments and a magnetic transition below 1.3 K. We shall see that this surprising behavior and other features of the experiment, such as the small amount of entropy lost at the transition, can be understood if we take into account the presence of a substantial non-cubic crystal field on the Ir sites. This crystal field arises from distortions of the oxygen octahedral that are evident from the crystal structure inferred from an X-ray analysis. Finally, we observe unusual metamagnetic behavior at low-temperatures whose origin lies in the face-centered cubic (FCC) lattice that the Ir moments form in this ordered double perovskite (see e.g. [22, 23, 25-29]). Quantum magnetism on the frustrated FCC lattice is yet poorly understood both experimentally and theoretically, and so our work calls for detailed neutron scattering and magnetic X-ray studies of this unusual quantum magnet.

**Table 1:** *Lattice Parameters for $Sr_2YIrO_6$ and $Sr_2GdIrO_6$*

| Compound | Structure | Space group | a (Å) | b (Å) | c (Å) | β (°) | $IrO_6$ |
|---|---|---|---|---|---|---|---|
| $Sr_2YIrO_6$ | Monoclinic | $P2_1/n$ | 5.7751 | 5.7919 | 8.1704 | 90.22 | Flattened |
| $Sr_2GdIrO_6$ | Cubic | Pn-3 | 8.2392 | | | | |

For contrast and comparison, we also present data for the double-perovskite $Sr_2GdIrO_6$, whose structure is much less distorted, as discussed in *Supplemental Material*. $Sr_2YIrO_6$ adopts a monoclinic structure essentially derived from the perovskite $SrIrO_3$ by replacing every other Ir by nonmagnetic Y; the remaining magnetic $Ir^{5+}$ ions form a network of edge-sharing tetrahedra or a FCC structure with lattice parameters elongated compared to the parent cubic structure, as shown in **Fig. 1**. Because of the differences in valence state and ionic radius between $Y^{3+}$ and $Ir^{5+}$ ions, no significant intersite disorder is expected. The lattice parameters of $Sr_2YIrO_6$ are given in **Table 1**. The $IrO_6$ octahedra are tilted and rotated, as seen in **Fig. 1**. A crucial structural detail is that each $IrO_6$ octahedron is significantly flattened since the bond distance between Ir and apical oxygen Ir-O3 (= 1.9366 Å) is considerably shorter than the in-plane Ir-O1 and Ir-O2 bond distances (= 1.9798 Å, 19723 Å, respectively), as shown in **Fig. 1b**. The flattening of the $IrO_6$ octahedra generates a non-cubic crystal field $\Delta$ that strongly competes with the spin-orbit interaction $\lambda_{so}$, as discussed below.

Single crystals of $Sr_2YIrO_6$ and $Sr_2GdIrO_6$ were grown using a self-flux method from off-stoichiometric quantities of $IrO_2$, $SrCO_3$ and $Y_2O_3$ or $Gd_2O_3$, as described elsewhere [30, 31]. The diffraction data were refined using the full-matrix, least-squares SHELX-97 program [24]. Experimental details are described in *Supplementary Material.*

$Sr_2YIrO_6$ displays paramagnetic behavior at temperatures above 1.5 K, as the magnetic susceptibility $\chi(T)$ follows the Curie-Weiss law for $50 < T < 300$ K, as shown in **Fig. 2a**. Data fits to the Curie-Weiss law over the range $50 < T < 300$ K yields an effective moment $\mu_{eff} = 0.91$ $\mu_B$/Ir and a Curie-Weiss temperature $\theta_{CW} = -229$ K. The value of $\mu_{eff}$ is considerably smaller than the value 2.83 $\mu_B$/Ir expected for a conventional $S = 1$ 5d-electron system. In fact, a reduced value of $\mu_{eff}$ is commonplace in iridates [**21, 22, 30-33**] because the strong SOI causes a partial cancellation of the spin and orbital contributions [**22**]. A strong antiferromagnetic exchange coupling might be inferred from the large magnitude of $\theta_{CW}$ (= -229 K); however, the absence of any magnetic ordering at $T > 1.5$ K indicates the existence of strong quantum fluctuations in $Sr_2YIrO_6$. A signature for long-range antiferromagnetic order is evident at a very low temperature, $T_N = 1.3$ K, as shown in **Fig. 2b**. The two temperature scales evident in the magnetic data yield a strikingly large frustration parameter, $|\theta_{CW}|/T_N = 176.2$.

The magnetic state undergoes a sharp metamagnetic transition at a critical field $H_C$ below $T_N$, as shown in **Fig. 2c.** The isothermal magnetization M(H) initially rises and then exhibits a plateau before a rapid jump at $H_C$, which occurs at an applied magnetic field $\mu_oH = 2.5$ T for $T = 0.5$ K and $\mu_oH = 5.3$ T for $T = 0.8$ K. The metamagnetic transition signals a spin reorientation; but the remarkably low ordered moment (< 0.023 $\mu_B$/Ir, even at $H > H_C$) implies that the magnetic state is only partially ordered or unsaturated. Note that a field dependence of M(H) that features a plateau followed by a metamagnetic transition is observed for some geometrically frustrated magnets [**34**].

The onset of long-range magnetic order is confirmed by an anomaly in the specific heat C(T) observed near $T_N$, as shown in **Fig. 3a**. This anomaly is well defined but weak, and is extremely sensitive to low magnetic fields; for example, the magnetic anomaly is considerably enhanced at $\mu_o H = 1$ T, but is suppressed by $\mu_o H = 7$ T, as can be seen in **Fig. 3b**. The entropy removal S(T) due to the magnetic transition is finite but very small compared to that expected for any possible magnetic ground states consistent with $J_{eff} = 1$ or 2, or S = 1.

In light of the data presented above, it is clear that neither SOC $\lambda$ nor non-cubic crystal field $\Delta$ alone dominates the low-temperature behavior of $Sr_2YIrO_6$. Indeed, for $\lambda \gg \Delta$, a prevailing SOC would suppress magnetic order and render a singlet ground state $J_{eff} = 0$ (Fig. 4 (c)), which is clearly inconsistent with the experimental observation. On the other hand, a S = 1 ground state would occur if $\Delta \gg \lambda$ (Fig. 4(b)). This scenario cannot adequately account for the small amount of entropy lost at the transition. It is therefore compelling to attribute the observed magnetic state to a delicate interplay between the competing $\lambda$ and $\Delta$. The standard accepted Hamiltonian for electrons in a $t_{2g}$ manifold is given by a sum of one-body terms: $\lambda$, $\Delta$, U and Hund's rule exchange $J_H$,

$$H_{t2g} = H_{1b} + H_{mb} \qquad (1)$$

$$H_{1b} = \sum_{m,m',s,s'} c^\dagger_{ms} (\lambda \vec{l}_{mm'} \cdot \vec{s}_{ss'} + \Delta (\vec{l} \cdot \hat{n})^2_{mm'} \delta_{ss'}) c_{m's'}$$

$$H_{mb} = U\left(\sum_{m,s} c^\dagger_{ms} c_{ms}\right)^2 + \frac{J_H}{2} \sum_{m,m',s,s'} \left(c^\dagger_{ms} c^\dagger_{m's'} c_{ms'} c_{m's} + c^\dagger_{ms} c^\dagger_{m's'} c_{m's'} c_{m's}\right) \quad (2)$$

where *m* is an index that labels the *yz; xz; xy* orbitals. $\vec{l}$ and $\vec{s}$ are the spin-1 and spin-1/2 Pauli matrices. $\hat{n}$ is a unit vector in the direction of the non-cubic distortion or an Ir-O bond. Since the $Ir^{5+}$ ion carries four 5d-electrons in the $t_{2g}$ orbitals, **Eq. (1)** becomes a 15 × 15 matrix in this space (see **Fig. 4(a)**). We ignore U as we compare the quantum states with the same number of particles, thus we are left with $\lambda$, $\Delta$ and $J_H$. Since the flattening of the $IrO_6$ octahedra (see **Fig. 1**) renders $\Delta < 0$ and both $\lambda$ and $J_H > 0$, the $Ir^{5+}$ ion always has a non-degenerate ground state independent of the magnitude of the couplings. However, the excitation gap to the lowest doublet becomes substantially suppressed in the regime where all three parameters are comparable, as shown in **Fig. 4**.

The physics of this regime can be understood by first diagonalizing the problem with $\Delta$ and $J_H$. In essence, the non-cubic crystal field $\Delta < 0$ splits the $t_{2g}$ orbitals, leaving the $a_{1g}$ having lower energy than the $e_g$ states. Populating the states with four 5d-electrons gives a degeneracy of 6 (3 singlets and 1 triplet); the presence of $J_H > 0$ then promotes a robust S = 1 ground state, as discussed above. Now *adding the SOC $\lambda$ to the interactions splits the triplet, resulting in a singlet ground state and a doublet excited state*. This is the near degeneracy, as shown in **Fig.4**. If $\Delta > \lambda$, the slitting, $\delta$, can be fairly small.

While a full super-exchange calculation of the interaction between Ir moments is possible, a cartoon model can capture the essence of the magnetism. Following our discussion above, we first consider the exchange interactions in the absence of $\lambda$, where spin rotation symmetry is preserved. We hence expect the resulting *S = 1* moments to interact with a Heisenberg interaction on the FCC lattice. Now, adding $\lambda$ to the interactions yields a local term $\delta (S_i^z)^2$ on the $i^{th}$ Ir site. Putting these together we arrive at the following S=1 model $H_p = J \sum_{ij} \vec{S}_i \cdot \vec{S}_j + \delta \sum_i (S_i^z)^2$. Such models have been studied extensively on bipartite lattices (although not on the FCC lattice of interest here) both in theory **[35]** and experiment **[36]**. The broad feature of such models that is important for our discussion here is that the magnetic order found when $\delta = 0$ can persist even when $\delta > 0$, albeit with a suppression of the magnetic ordering temperature. As the coupling $\delta$ is further increased, at critical value of the ratio $(\delta/J)_c$, the magnetic order is completely suppressed at a quantum critical point.

This scenario predicts that a magnetic ordering can occur even though a single ion can be in a ***non-degenerate "singlet" state***. A unique characteristic of such a magnetic order is that the entropy, which is removed at the magnetic transition, is much smaller than the Rln(3) = 9.13 J/mole K expected for an ordering transition of **S** = 1 moments as the isolated ions already lose their entropy when T << $\delta$. This prediction is consistent with the experimental observation that the entropy is only ~ 0.01 J/mole K, a tiny fraction of Rln(3), as shown in **Fig. 3c**. Indeed, the ground state is so fragile that even low magnetic fields are strong enough to tip the balance, based upon the data presented herein; this qualitatively explains **(a)** the strongly depressed, weak magnetic order (**Fig. 2b**), **(b)** the drastic changes in the specific heat and entropy in weak applied magnetic fields (**Fig. 3**), and **(c)** the low-field metamagnetic transition (**Fig. 2c**).

In conclusion, the relative effect of the SOC critically depends upon the strength of non-cubic crystal fields, electron hopping and exchange interactions; therefore it should vary from material to material. While the $J_{eff}$ model successfully captures the new physics observed in many iridates, it may not be appropriate to describe new phenomena in other heavy transition metal materials with ***strong non-cubic crystal fields***. Our results for $Sr_2YIrO_6$ not only illustrate a prototypical breakdown of spin-orbit-driven $J_{eff}$ states, but also provides a new paradigm for studying novel phenomena emerging from the strong competition between the SOC and crystal fields.

**Acknowledgments**

GC is very thankful to Drs. D. Khomskii, T.Saha-Dasgupta, G. Jackeli, Y. B. Kim, D. Singh and S. Stroltsov for enlightening discussions. This work was supported by the US National Science Foundation via grants DMR-0856234 , DMR-1265162 and DMR-NSF DMR-1056536 (RKK).

**References:**

1. B. J. Kim, Hosub Jin, S. J. Moon, J.-Y. Kim, B.-G. Park, C. S. Leem, Jaejun Yu, T. W. Noh, C. Kim, S.-J. Oh, V. Durairai, G. Cao & J.-H. Park *Phys. Rev. Lett.* 101, 076402 (2008)

2. S.J. Moon, H. Jin, K.W. Kim, W.S. Choi, Y.S. Lee, J. Yu, G. Cao, A. Sumi, H. Funakubo, C. Bernhard, and T.W. Noh *Phys. Rev. Lett.* 101, 226401 (2008)

3. B. J. Kim, H. Ohsumi, T. Komesu, S. Sakai, T. Morita, H. Takagi and T. Arima, *Science* **323**, 1329 (2009)

4. Fa Wang and T. Senthil, *Phys. Rev. Lett.* **106**, 136402 (2011)

5. Xiangang Wan, A. M. Turner, A. Vishwanath and S. Y. Savrasov *Phys. Rev. B* **83**, 205101 (2011)

6. G. Jackeli, and G. Khaliullin, Phys. Rev. Lett. **102**, 017205 (2009)

7. G.-W. Chern and N. B. Perkin, Phys. Rev. B 80, 180409 (R) (2009)

8. J. Chaloupka, G. Jackeli, and G. Khaliullin, Phys. Rev. Lett. **105**, 027204 (2010)

9. A. Shitade, H. Katsura, J. Kuneš, X.-L. Qi, S.-C. Zhang, and N. Nagaosa, Phys. Rev. Lett. **102**, 256403 (2009)

10. Y. Singh and P. Gegenwart, Phys. Rev. B **82**, 064412 (2010)

11. X. Liu, T. Berlijn, W. G. Yin, W. Ku, A. Tsvelik, Y.-J. Kim, H. Gretarsson, Y. Singh, P. Gegenwart, and J. P. Hill, Phys. Rev. B **83**, 220403 (2011).

12. S. K. Choi*,* R. Coldea, A. N. Kolmogorov, T. Lancaster, I.I. Mazin, S. J. Blundell, P. G. Radaelli, Yogesh Singh, P. Gegenwart, K.R. Choi, S.-W. Cheong, P. J. Baker, C. Stock, and J. Taylor, Phys. Rev. Lett. **108**, 127204 (2012)

13. F. Ye, S. Chi, H. Cao, B. C. Chakoumakos, J. A. Fernandez-Baca, R. Custelcean, T. F. Qi, O. B. Korneta, and G. Cao, Phys. Rev. B **85**, 180403 (R) (2012)

14. Crag Price and Natalia B. Perkins, Phys. Rev. Lett. **109**, 187201 (2012)

15. J. Chaloupka, G. Jackeli, and G. Khaliullin, Phys. Rev. Lett. **110**, 97204 (2013)

16. C. H. Kim, H. S. Kim, H. Jeong, H. Jin, and J. Yu, Phys. Rev. Lett. **108**, 106401 (2012)

17. S. Bhattacharjee, S. S. Lee, and Y. B. Kim, New J. of Phys. **14**, 073015 (2012)

18. I. I. Mazin, H. O. Jeschke, K. Foyevtsova, R. Valentí, and D. I. Khomskii, Phys. Rev. Lett. **109**, 197201 (2012)

19. M. Bremholm, S.E.Dutton, P.W. Stephens, R. J.Cava, J Solid State Chem., **184**, 601 (2011)

20. X. Liu, Vamshi M. Katukuri, L. Hozoi, Wei-Guo Yin, M. P. M. Dean, M. H. Upton, Jungho Kim, D. Casa, A. Said, T. Gog, T. F. Qi, G. Cao, A. M. Tsvelik, Jeroen van den Brink, and J. P. Hill, Phys. Rev. Lett. **109**, 157401 (2012)

21. H. Gretarsson, J. P. Clancy, X. Liu, J. P. Hill, E. Bozin, Y. Singh, S. Manni, P. Gegenwart, J. Kim, A. H. Said, et al., Phys. Rev. Lett. **110**, 076402 (2013)

22. Gang Chen, Rodrigo Pereira and Leon Balents, Phys. Rev. B **82**, 174440 (2010)

23. A. S. Erickson, S. Misra, G. J. Miller, R. R. Gupta, Z. Schlesinger, W. A. Harrison, J. M. Kim, and I. R. Fisher, Phys. Rev. Lett. **99**, 016404 (2007)

24. G. M. Sheldrick, *Acta Crystallogr* A **64**, 112 (2008)

25. T. Aharen, J. E. Greedan, C. A. Bridges, A. A. Aczel, J. Rodriguez, G. MacDougall, G. M. Luke, T. Imai, V. K. Michaelis, S. Kroeker, H. Zhou, C. R. Wiebe and L. M. D. Cranswick, Phys. Rev. B **81**, 224409 (2010)

26. M. A. de Vries, A. C. Mclaughlin, and J.-W. G. Bos, Phys. Rev. Lett. **104**, 177202 (2010)

27. O. Erten, O. N. Meetei, A. Mukherjee, M. Randeria, N. Trivedi, and P. Woodward, Phys. Rev. Lett. **107**, 257201 (2011)

28. Santu Baidya and T. Saha-Dasgupta, private communications, 2013

29. P. M. Woodward, J. Goldberger, M. W. Stoltzfus, H. W. Eng, R. A. Ricciardo, P. N. Santhosh, P. Karen, A. R. Moodenbaugh, *J. Amer. Ceram. Soc.* **91**, 1796-1806 (2008)

30. M. Ge, T. F. Qi, O.B. Korneta, L.E. De Long, P. Schlottmann and G. Cao, Phys. Rev. B **84**, 100402(**R**) (2011)

31. S. Chikara, O. Korneta, W. P. Crummett, L. E. DeLong, P. Schlottmann and G. Cao *Phys. Rev. B* **80**, 140407 (**R**) (2009)

32. Feng Ye, Songxue Chi, Bryan C. Chakoumakos, Jaime A. Fernandez-Baca, Tongfei Qi, and G. Cao, *Phys. Rev. B* **87**, 140406(R) (2013)

33. D. Haskel, G. Fabbris, Mikhail Zhernenkov, M. van Veenendaal, P. Kong, C. Jin, G. Cao, *Phys. Rev Lett.* **109**, 027204 (2012)

34. G. Cao, V. Durairaj, S. Chikara, S. Parkin and P. Schlottmann, *Phys. Rev. B* 75, 134402 (2007)

35. Han-Ting Wang and Yupeng Wang, Phys. Rev. B **71**, 104429 (2005)

36. S. A. Zvyagin, J. Wosnitza, C. D. Batista, M. Tsukamoto, N. Kawashima, J. Krzystek, V. S. Zapf, M. Jaime, N. F. Oliveira, Jr., and A. Paduan-Filho. Phys. Rev. Lett. **98**, 047205 (2007)

**Captions:**

**Fig. 1. (a)** The double perovskite crystal structure of $Sr_2YIrO_6$ based on the single-crystal diffraction data; **(b)** Enlarged view of flattened octahedra $IrO_6$; **(c)** The ordered replacement of nonmagnetic Y ions for magnetic Ir ions leading to a face centered cubic (FCC) lattice with geometrically frustrated edge-sharing tetrahedra formed by the pentavalent $Ir^{5+}$ ions in $Sr_2YIrO_6$. Note that octahedra $IrO_6$ are noticeably tilted and rotated.

**Fig. 2.** Magnetic properties of $Sr_2YIrO_6$: The temperature dependence of **(a)** the magnetic susceptibility $\chi(T)$ (left scale) and $1/\Delta\chi$ (right scale) at $\mu_oH = 0.1$ T for $1.7$ K $\leq T < 350$ K (Note that $\Delta\chi=\chi-\chi_o$, where $\chi_o$ is a temperature-independent contribution to $\chi$), and **(b)** the low-temperature magnetization M(T) at $\mu_oH = 7$ T; **(c)** The isothermal magnetization M(H) at T=0.5, 0.8 and 1.7 K.

**Fig. 3.** Thermal properties of $Sr_2YIrO_6$: For $0.05$ K $\leq T \leq 5$ K, the temperature dependence of **(a)** the specific heat C(T) at zero field $\mu_oH = 0$ T, **(b)** C(T) at $\mu_oH = 0, 1, 3,$ and 7 T, and **(c)** the entropy S(T) at $\mu_oH = 0, 1, 3,$ and 7 T.

**Fig. 4.** Ground state of a single $Ir^{5+}$ ($5d^4$) ion under three different scenarios. **(a)** When $\Delta$, $J_H$ and $\lambda=0$, the four electrons in the $t_{2g}$ give 15 ground states. Throughout our analysis here we use the physical $J_H>0$, $\Delta<0$ and $\lambda>0$. **(b)** When $\lambda<<\Delta$, the ground state is a S=1 triplet. **(c)** When $\lambda>>\Delta$, the ground state is thought of as a $J_{eff}=0$ singlet. The main inset shows the spectrum of 4 electrons $E_4$ as a function of $\lambda$ for some expected values $J_H =-\Delta=0.5$ eV. At intermediate physical values of $\lambda$ the ground state is always a singlet but with a low-lying doublet. *Note:* The doublets are non-Kramers time reversed pairs and the line originating at $E_4=2$ has both a doublet and singlet that are not exactly degenerate but have a splitting too small to see on the scale here.

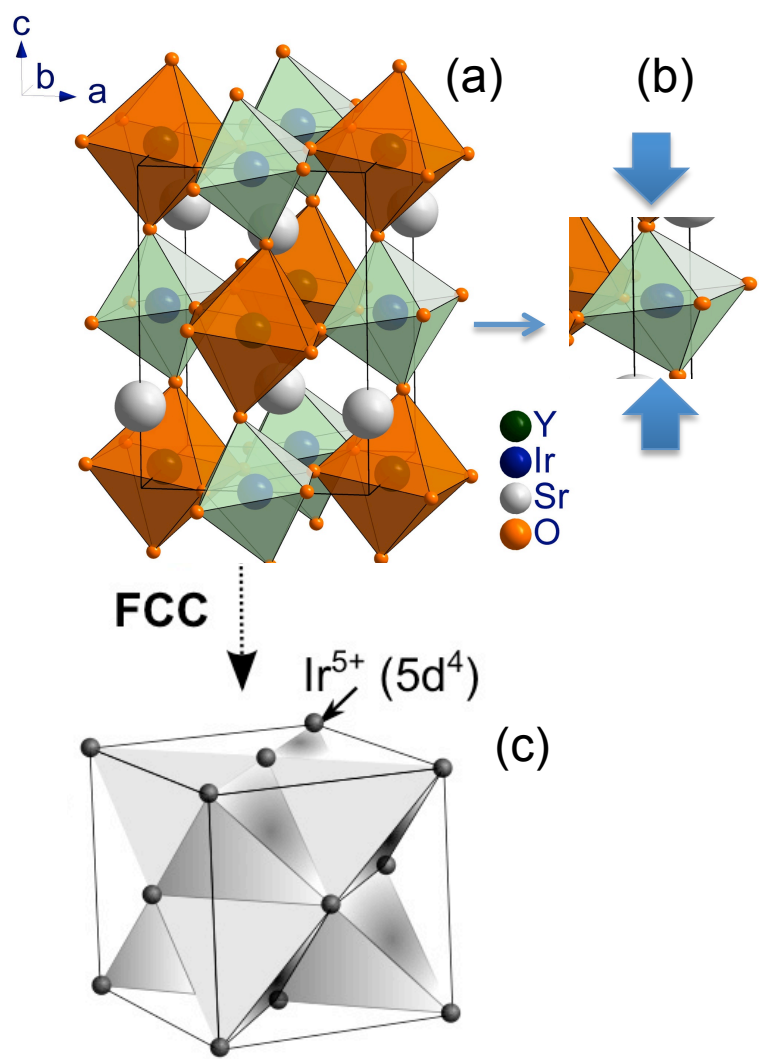

Fig. 1

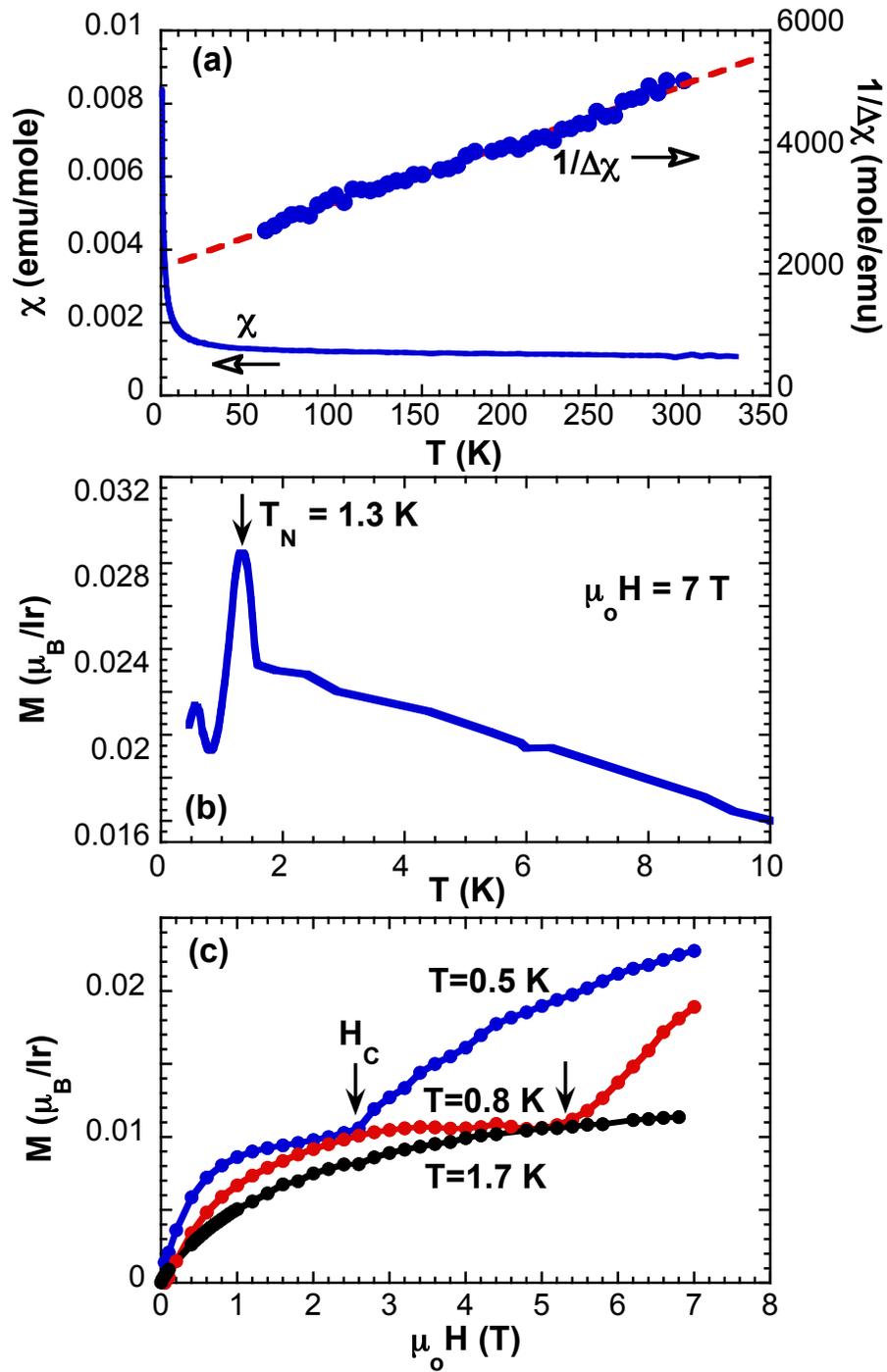

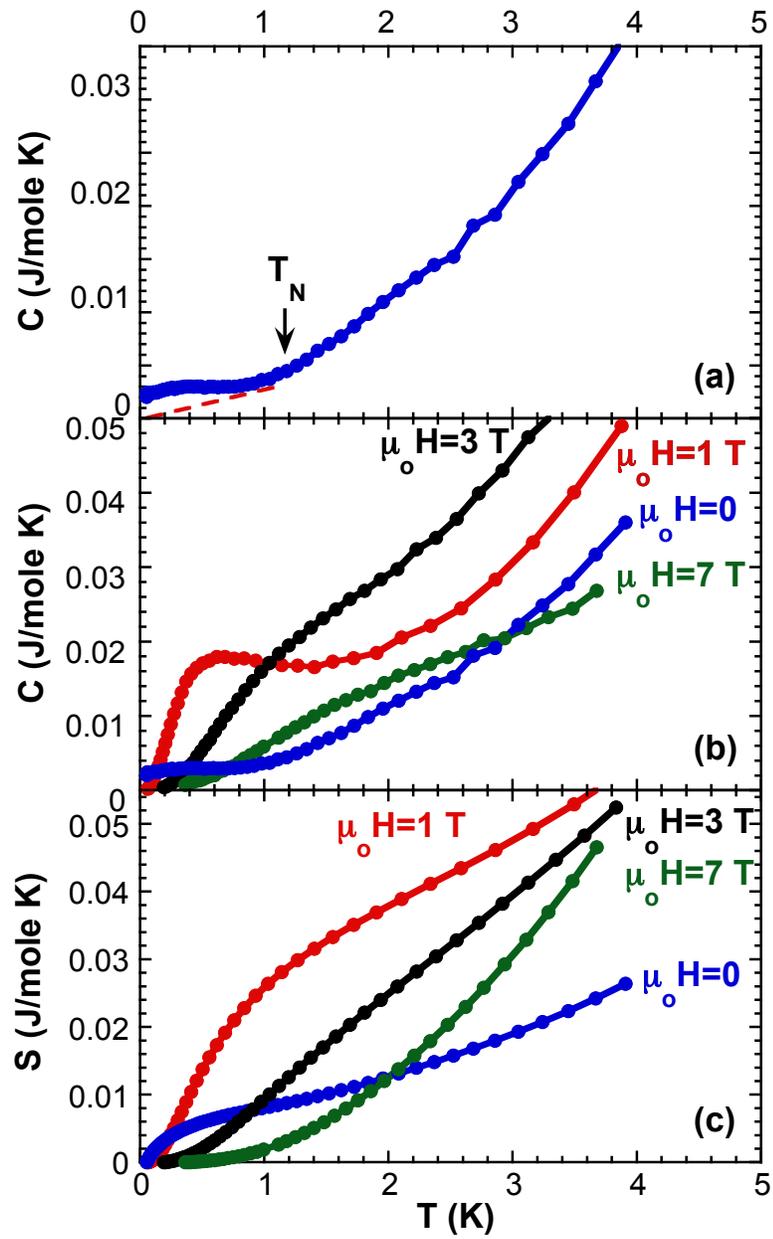

Fig. 3

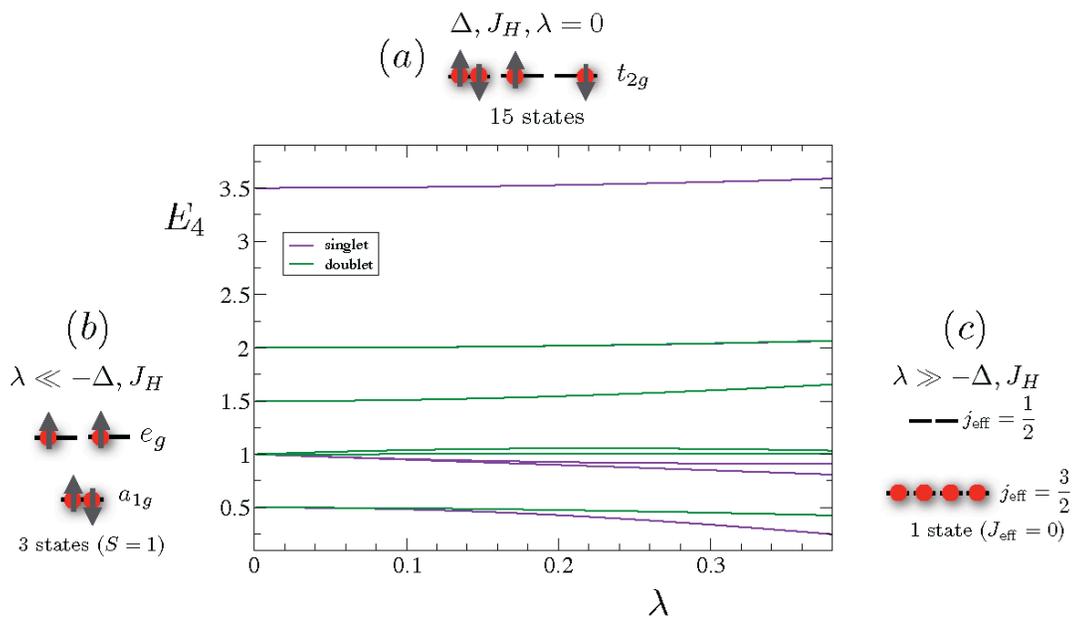

Fig. 4